\documentclass[12pt]{article} %
\usepackage{graphicx,amssymb,amsmath,mathtools,bm} 
\usepackage{bm,url}%

\date{} %

\begin{document}

\title{A constructive approach to simulating quantum phenomena by plasma-like collections}

\author{
       Andrey Akhmeteli \\ 
        LTASolid Inc. \\ akhmeteli@ltasolid.com
       }%

\maketitle

\thispagestyle{empty}


\begin{abstract}
Previously, the author offered a plasma-like description of quantum phenomena. This article offers a new criterion of approximation of probability density functions of quantum theories by sums of $\delta$-functions with integer coefficients and a constructive approach to building such sets of $\delta$-functions.
\end{abstract}
Schrödinger noticed in 1952 ~\cite{Schroed} that a scalar complex wave function can be made real by a gauge transformation. It turned out that one real function is also enough to describe matter in more realistic theories, such as the Dirac equation in an arbitrary electromagnetic  ~\cite{Akhmeteli-JMP,Akhmeteli-EPJC2} or Yang-Mills ~\cite{Akhmeteliqr} field. As these results suggest some ``symmetry'' between positive and negative frequencies and, therefore, particles and antiparticles, one-particle wave functions can be described as plasma-like collections of a large number of particles and antiparticles ~\cite{Akhmeteli-EPJC,Akhm-Ent,Akhmeteliqr}. The similarity of the dispersion relations for the Klein-Gordon equation and a simple plasma model provides another motivation for the plasma-like description of quantum particles.

The criterion for approximation of continuous charge density distributions by discrete ones with quantized charge is the equality of partial Fourier sums. It is proven for the one-dimensional case ~\cite{Akhmeteliqr} and later for arbitrary dimensions ~\cite{fedja} that such approximation can be arbitrarily precise as defined by this criterion.

The wonderful proof of ~\cite{fedja} is important as it shows that plasma-like collections of ~\cite{Akhm-Ent, Akhmeteliqr} can emulate any experimental effects described by, say, the Dirac equation. However, the proof is not (quite) constructive. For example, it uses Brouwer's fixed point theorem. A constructive approach is desirable as it would enable building plasma-like collections approximating arbitrary wave functions. Such collections can be used to emulate experiments described by quantum theory, such as the double-slit experiment.

To provide a constructive approach inspired by ~\cite{fedja} (and also by ~\cite{Koures}), a modified criterion of proximity of a smooth probability density and a collection of $\delta$-functions with integer coefficients is used in this work. This criterion is mathematically different from the previous one but seems also adequate for applications to quantum physics.  

The previous criterion was as follows.

Let us consider a smooth real function $f(\boldsymbol{x})$ in a $d$-dimensional cube $C:-\pi\leq x^i\leq \pi$, where $\boldsymbol{x}=(x^1,\ldots,x^d)$, $\int\limits_C f(\boldsymbol{x})\boldsymbol{dx}=1$, and $f(\boldsymbol{x})$ vanishes on the boundary of the cube.

Let us also consider distributions of the form
$g(\boldsymbol{x})=\delta(\boldsymbol{x}-\boldsymbol{x}_0)+\sum\limits_{i=1}^{n}(\delta(\boldsymbol{x}-\boldsymbol{y}_i)-\delta(\boldsymbol{x}-\boldsymbol{z}_i))$,
where $\delta(\boldsymbol{x})$ is the $d$-dimensional Dirac's $\delta$-function, and $\boldsymbol{x}_0,\boldsymbol{y}_i,\boldsymbol{z}_i$ belong to the interior of $C$.

Functions $f(\boldsymbol{x})$ and $g(\boldsymbol{x})$ in the cube can be expanded in Fourier series of the form
\begin{equation}\label{eq:dig1}
f(\boldsymbol{x})=\sum\limits_{\boldsymbol{k} \in\mathbb{Z}^d}f_{\boldsymbol{k}}\mathrm{e}^{i\boldsymbol{k}\boldsymbol{x}},
\end{equation}
 where $d$-dimensional vectors $\boldsymbol{k}=(k^1,\ldots,k^d)$ have integer coordinates, $f_{\boldsymbol{k}}=f_{\boldsymbol{-k}}^*$ for real functions, and
\begin{equation}\label{eq:dig2}
f_{\boldsymbol{k}}=(2\pi)^{-d}\int\limits_{C}f(\boldsymbol{x})\mathrm{e}^{-i\boldsymbol{k}\boldsymbol{x}}{\boldsymbol{d}\boldsymbol{x}},
\end{equation}
For some natural $m$, the series can be truncated by keeping only terms with such $\boldsymbol{k}$ that $|k^i|\leq m$:
\begin{equation}\label{eq:dig1tr}
f_{tr}(\boldsymbol{x})=\sum_{\mathclap{\substack{\boldsymbol{k} \in\mathbb{Z}^d\\ |k^i|\leq m}}}f_{\boldsymbol{k}}\mathrm{e}^{i\boldsymbol{k}\boldsymbol{x}}.
\end{equation}
Coincidence of the truncated series for $f(\boldsymbol{x})$ and $g(\boldsymbol{x})$ for some $m$ was used as the criterion of proximity. The following statement was proven in ~\cite{Akhmeteliqr} (for $d=1$) and in ~\cite{fedja} (for an arbitrary $d)$:

$\forall m$ $\exists\,n,\boldsymbol{x}_0,\boldsymbol{y}_i,\boldsymbol{z}_i$ ($1\leq i\leq n$) such that the truncated Fourier series for $f(\boldsymbol{x})$ and $g(\boldsymbol{x})$ coincide.

The new criterion requires that $f(\boldsymbol{x})$ and $g(\boldsymbol{x})$ become close after Gaussian smoothing
\begin{equation}\label{eq:dig1gs}
g_{\delta}(\boldsymbol{x})=\int g(\boldsymbol{\xi})G_{\delta}({\boldsymbol{x}-\boldsymbol{\xi})\boldsymbol{d}\boldsymbol{\xi}},
\end{equation}
 a convolution with the Gaussian
\begin{equation}\label{eq:dig1ga}
G_{\delta}(\boldsymbol{x})=(\sqrt{2 \pi} \delta)^{-d} e^{-\frac{\boldsymbol{x}^2}{2\delta^2}}.
\end{equation} 
We do not require anymore that the smooth function $f(\boldsymbol{x})$ be only defined in the $d$-dimensional cube $C$.
Then the modified statement (for the function $f(\boldsymbol{x})$ satisfying the conditions above) is as follows.

$\forall \delta>0,\varepsilon>0$ $\exists\,n,\boldsymbol{x}_0,\boldsymbol{y}_i,\boldsymbol{z}_i$ ($1\leq i\leq n$) such that $|f_{\delta}(\boldsymbol{x})-g_{\delta}(\boldsymbol{x})|<\varepsilon$.

A sketch of the proof. may be as follows. Let us define the following function:
\begin{equation}\label{eq:ga1}
\rho(\boldsymbol{x})=G_{\delta_n}(\boldsymbol{x-x_0})-f(\boldsymbol{x}),
\end{equation}
where $x_0$ is chosen arbitrarily and $\delta_n\ll\delta$. One can see that
\begin{equation}\label{eq:ga2}
\int\boldsymbol{dx}\rho(\boldsymbol{x})=0.
\end{equation}
Let $\varphi(\boldsymbol{x})$ be a solution of the Poisson equation
\begin{equation}\label{eq:ga3}
\bigtriangleup\varphi(\boldsymbol{x})=-\rho(\boldsymbol{x}),
\end{equation}
such that $\varphi(\boldsymbol{x})=0$ at infinity. Using Green's first identity with a vanishing ``surface'' integral, we obtain
\begin{eqnarray}\label{eq:ga4}
\nonumber
\int\bm{\nabla}_{\boldsymbol{x'}}G_{\delta}(\boldsymbol{x'}-\boldsymbol{x})\bm{\nabla}_{\boldsymbol{x'}}\varphi(\boldsymbol{x'})\boldsymbol{dx'}=-\int G_{\delta}(\boldsymbol{x'}-\boldsymbol{x})\bigtriangleup_{\boldsymbol{x'}}\varphi(\boldsymbol{x'})\boldsymbol{dx'}=\\
-\int G_{\delta}(\boldsymbol{x'}-\boldsymbol{x})\rho(\boldsymbol{x'})\boldsymbol{dx'},
\end{eqnarray}
On the other hand, for  $a\ll\delta_n$ (we also have $a\ll\delta$ in this case), $\boldsymbol{l}=(l_1,\ldots,l_j,\ldots,l_d)$, where $l_j$ are integer and $-\infty\leq l_j\leq \infty, \boldsymbol{x_l}=a\boldsymbol{l}$, we obtain, using a simple cubature formula and a slightly modified definition of the directional derivative as a limit:
\begin{eqnarray}\label{eq:ga5}
\nonumber
\int\bm{\nabla}_{\boldsymbol{x'}}G_{\delta}(\boldsymbol{x'}-\boldsymbol{x})\bm{\nabla}_{\boldsymbol{x'}}\varphi(\boldsymbol{x'})\boldsymbol{dx'}\approx a^d\sum_{\mathclap{\substack{l^j=-\infty\\ 1\leq j\leq d}}}^{\infty}\bm{\nabla}_{\boldsymbol{x_l}}G_{\delta}(\boldsymbol{x_l}-\boldsymbol{x})\bm{\nabla}_{\boldsymbol{x_l}}\varphi(\boldsymbol{x_l}) \approx
\\
\nonumber
a^d\sum_{\mathclap{\substack{l^j=-\infty\\ 1\leq j\leq d}}}^{\infty}\frac{G_{\delta}\left(\boldsymbol{x_l}+\frac{1}{2}a^d\bm{\nabla}_{\boldsymbol{x_l}}\varphi(\boldsymbol{x_l})-\boldsymbol{x}\right)-G_{\delta}\left(\boldsymbol{x_l}-\frac{1}{2}a^d\bm{\nabla}_{\boldsymbol{x_l}}\varphi(\boldsymbol{x_l})-\boldsymbol{x}\right)}{a^d}=
\\
\sum_{\mathclap{\substack{l^j=-\infty\\ 1\leq j\leq d}}}^{\infty}\left(G_{\delta}\left(\boldsymbol{x_l}+\frac{1}{2}a^d\bm{\nabla}_{\boldsymbol{x_l}}\varphi(\boldsymbol{x_l})-\boldsymbol{x}\right)-G_{\delta}\left(\boldsymbol{x_l}-\frac{1}{2}a^d\bm{\nabla}_{\boldsymbol{x_l}}\varphi(\boldsymbol{x_l})-\boldsymbol{x}\right)\right).
\end{eqnarray}
We replaced the derivatives with difference quotients. This approximation gets better with smaller $a$.

One can see from  (\ref{eq:ga1}), (\ref{eq:ga4}), and  (\ref{eq:ga5}) that
\begin{eqnarray}\label{eq:ga6}
\nonumber
f_{\delta}(\boldsymbol{x})\approx G_{\delta}(\boldsymbol{x-x_0})+
\\
\sum_{\mathclap{\substack{l^j=-\infty\\ 1\leq j\leq d}}}^{\infty}\left(G_{\delta}\left(\boldsymbol{x_l}+\frac{1}{2}a^d\bm{\nabla}_{\boldsymbol{x_l}}\varphi(\boldsymbol{x_l})-\boldsymbol{x}\right)-G_{\delta}\left(\boldsymbol{x_l}-\frac{1}{2}a^d\bm{\nabla}_{\boldsymbol{x_l}}\varphi(\boldsymbol{x_l})-\boldsymbol{x}\right)\right)
\end{eqnarray}
(we used the condition $\delta_n\ll\delta$ to replace the result of Gaussian smoothing of function $G_{\delta_n}(\boldsymbol{x-x_0})$ with $G_{\delta}(\boldsymbol{x-x_0})$. Note that a convolution of Gaussian functions with standard deviations $\delta_1$ and $\delta_2$ is another Gaussian function with standard deviation $\sqrt{\delta_1^2+\delta_2^2}$ ~\cite{Vinga}). One can have an arbitrarily good approximation leaving only a finite number of "dipoles" in the sum in the right-hand side of (\ref{eq:ga6}) due to (\ref{eq:ga2}) and (\ref{eq:ga3}).

Thus, probability density functions built from one-particle wave functions of quantum theory can be approximated arbitrarily well by a set of $\delta$-functions with integer coefficients. A constructive approach to building such a set was developed. This allows simulation of quantum effects using plasma-like collections. Let us also note that the approach implies an analogy with vacuum polarization.

\bibliographystyle{unsrt}

\end{document}